\documentstyle[12pt,equation]{article}
\begin{document}
\def\pslt{p\llap/_T}
\def\eslt{E\llap/_T}
\def\to{\rightarrow}
\def\te{\tilde e}
\def\tt{\tilde t}
\def\tb{\tilde b}
\def\ttau{\tilde \tau}
\def\tg{\tilde g}
\def\tga{\tilde \gamma}
\def\tnu{\tilde\nu}
\def\tl{\tilde\ell}
\def\tq{\tilde q}
\def\tw{\widetilde W}
\def\tz{\widetilde Z}
\def\tx{\widetilde \chi}
\newcommand{\cleqn}{\setcounter{equation}{0}}
\newcommand{\mst}{\mbox{$m_{\tilde{t}_1}$}}
\newcommand{\mgl}{\mbox{$m_{\tilde{g}}$}}
\newcommand{\mhalf}{\mbox{$m_{1/2}$}}
\newcommand{\mzino}{\mbox{$m_{\tilde{Z}_1}$}}
\newcommand{\mx}{\mbox{$M_X$}}
\newcommand{\brac}{\left( \frac{m_t}{190 \ {\rm GeV} \ \sin \! \beta}
                   \right)}
\newcommand{\mz}{\mbox{$M_Z$}}
\newcommand{\mt}{\mbox{$m_t$}}
\newcommand{\mb}{\mbox{$m_b$}}
\newcommand{\eplem}{\mbox{$e^+e^-$}}
\newcommand{\mpl}{\mbox{$M_P$}}
\newcommand{\sym}{\mbox{$SU(2) \times U(1)_Y$}}
\newcommand{\epem}{\mbox{$e^+e^-$}}
\newcommand{\tanb}{\mbox{$\tan\!\beta$}}
\newcommand{\bbbar}{\mbox{$b \bar{b}$}}
\newcommand{\lsp}{\mbox{$\tilde{Z}_1$}}
\newcommand{\st}{\mbox{$\tilde{t}_1$}}
\newcommand{\sbo}{\mbox{$\tilde{b}$}}
\newcommand{\stau}{\mbox{$\tilde{\tau}$}}
\newcommand{\be}{\begin{equation}}
\newcommand{\ee}{\end{equation}}
\newcommand{\een}{\end{subequations}}
\newcommand{\ben}{\begin{subequations}}
\newcommand{\beq}{\begin{eqalignno}}
\newcommand{\eeq}{\end{eqalignno}}
\renewcommand{\thefootnote}{\fnsymbol{footnote} }
\begin{flushright}
MAD/PH/843\\
July 1994
\end{flushright}
\vspace{1.5cm}
\begin{center}
{\Large \bf Phenomenology of the Minimal Supergravity $SU(5)$
Model\footnote{Based on a talk presented at the {\it 3rd Workshop on High
Energy Particle Physics}, Madras, India, January 1994.}}\\
\vspace{5mm}
Manuel Drees\footnote{Heisenberg Fellow}\\
{\em Physics Department, University of Wisconsin, Madison, WI 53706, USA}
\end{center}

\begin{abstract}
The minimal grand unified supergravity model is discussed. Requiring radiative
breaking of the electroweak gauge symmetry, the unification of $b$ and $\tau$
Yukawa couplings, a sufficiently stable nucleon, and not too large a relic
density of neutralinos produced in the Big Bang constrains the parameter
space significantly. In particular, the soft breaking parameter \mhalf\ has to
be less than about 130 GeV, and the top quark Yukawa coupling has to be near
its (quasi) fixed point. The former condition implies $\mgl \leq 400$ GeV and
hence very large production rates for gluino pairs at the LHC, while the
latter constraint implies that the lighter stop and sbottom eigenstates are
significantly lighter than the other squarks, leading to characteristic
signatures for gluino pair events.

\end{abstract}
\clearpage
\pagestyle{plain}
\setcounter{page}{1}
\setcounter{footnote}{0}
\section*{1) Introduction}
Supersymmetry (SUSY) \cite{1} now seems to be the most popular extension of
the Standard Model (SM). There are several reasons for this. First of all,
SUSY solves the (technical) hierarchy problem \cite{2} (also known as
finetuning or naturalness problem), i.e. stabilizes the weak scale against
radiative corrections that otherwise tend to pull it up towards the GUT or
Planck scale. This is also true for SUSY's main competitor, technicolor (TC)
\cite{3}, although through a completely different mechanism. However, it seems
increasingly difficult to find realisations of the TC idea that are not ruled
out, or at least strongly disfavoured, by LEP measurements, in particular of
the so--called $S$ parameter \cite{4} and of the $Z \rightarrow b \bar{b}$
partial width \cite{5}. In contrast, sparticles decouple quickly, i.e. do not
affect predictions for LEP observables noticably if sparticle masses exceed
100 GeV or so, so that LEP measurements can only rule out SUSY if they also
exclude the SM (with a light Higgs boson) at the same time. Moreover, if
sparticles are light, agreement between LEP measurements and predictions is
often (slightly) improved compared to the non--supersymmetric SM \cite{6}.

Another strong motivation for SUSY is that the minimal supersymmetric standard
model (MSSM) allows for a predictive grand unification of all gauge
interactions, in contrast to the SM \cite{7}.\footnote{``Predictive" here
means that grand unification can be achieved for experimentally acceptable
values of the gauge coupling constants {\em without} having to resort to the
ad--hoc introduction of new fields with masses anywhere between the weak and
the GUT scale. Of course, SUSY does predict the existence of many new
(s)particles, but this is an inescapable consequence of the assumed symmetry.}
Actually GUTs were in better agreement with data \cite{8} with than without
SUSY already before LEP was turned on, but the higher precision achieved by
LEP experiments has made this argument much more convincing.

All these arguments are independent of the way SUSY is broken, provided only
that the breaking is ``soft" \cite{9} and sparticle masses do not (greatly)
exceed 1 TeV; if either of these conditions were violated, the naturalness
problem would re--appear. Unfortunately no completely convincing dynamical
mechanism of SUSY breaking has yet been suggested. For example, the one thing
we know for sure about SUSY breaking in superstring theories is that it does
not happen at any order in perturbation theory \cite{10}, i.e. SUSY breaking
is an intrinsically nonperturbative problem and thus not easily
treatable.\footnote{That should not be held against SUSY per se. Remember that
another prominent nonperturbative problem of modern field theory, the
confinement problem of QCD, has still not been solved, even though it only
involves fields with known mass and known interactions, and even though a
wealth of experimental data exists.} At present we are therefore forced to use
a phenomenological approach to SUSY breaking.

In many analyses of SUSY signals at colliders \cite{1,11} or elsewhere, it has
been assumed that the sparticle spectrum has a high degree of degeneracy at
the {\em weak} energy scale, ${\cal O}(100)$ GeV, where present and
near--future experiments operate. Specifically, it has often been assumed that
all squarks (with the possible exception of stops) are exactly degenerate in
mass with each other as well as with all sleptons. On the other hand, in these
same analyses the parameters of the Higgs sector have usually been chosen ``by
hand", {\em independently} of the sparticle spectrum. Both these assumptions
are, in my view at least, rather unnatural. First of all, we know
experimentally that within the MSSM the running gauge couplings meet (unify)
at $\mx \simeq 2 \cdot 10^{16}$ GeV. In other words, starting from a seemingly
complicated situation (described by three ``independent" gauge couplings) at
low energies we are led to a much simpler scenario with only a single gauge
coupling at very high energies. On the other hand, if we similarly run the
soft breaking parameters from the weak to the GUT scale, starting from a
degenerate spectrum at the weak scale leads to a complicated, highly
non--degenerate spectrum at the GUT scale. This violates what can be called
the ``unification dogma", which stipulates that nature should become simpler,
i.e. more symmetric, at higher energies. Secondly, the main beauty of SUSY is
that it naturally includes (elementary) scalar particles, which seem to be
required for the breaking of the electroweak symmetry in accordance with LEP
data. Some of this beauty is lost if we treat matter (sfermion) and Higgs
scalars differently.

It thus seems much more natural to me to assume a highly degenerate
(s)particle spectrum at the GUT (or even Planck) scale, and to extend this
degeneracy to include the Higgs bosons. This is called the minimal
supergravity (mSUGRA) scenario \cite{12}, the idea being that local
supersymmetry or supergravity is spontaneously broken in a ``hidden sector",
and that this is communicated to the visible (gauge/Higgs/matter) sector only
through flavour--blind gravitational--strength interactions. One very
attractive feature of this scenario is that it (almost) {\em automatically}
leads to the correct symmetry breaking pattern; that is, even though all
scalars get the same nonsupersymmetric (positive) mass term at scale \mx, at
the weak scale the Higgs fields (and {\em only} the Higgs fields) acquire
nonvanishing vacuum expectation values (vevs), provided only that the top
quark is not much lighter than the $W$ boson, which we now know to be true
\cite{13}. This ``miracle" occurs since radiative corrections involving
Yukawa interactions reduce the squared masses of the Higgs bosons, eventually
driving a combination of these masses to negative values \cite{12,14}, at
which point the electroweak gauge symmetry is broken. Notice that this
mechanism of radiative gauge symmetry breaking ``explains" both the existence
of the gauge hierarchy and the large mass of the top quark, in the sense that
it only works if $\log \left( \mx / \mz \right) \gg 1$ and the top Yukawa
$h_t \sim {\cal O}(1)$.

Of course, this approach to SUSY breaking is most naturally combined with
grand unification of all gauge interactions. Here I only consider the simplest
GUT group, $SU(5)$. Moreover, only the minimal necessary number of fields will
be assumed to exist at the GUT scale as well as at lower energies. As
discussed in more detail in sec. 2, this leads to a fairly predictive
(constrained) scenario \cite{15}, although some freedom in the choice of
parameters, and of the resulting phenomenology, still exists. It should be
emphasized that it is not at all trivial that this simplest of all SUSY GUTs
is still experimentally viable.\footnote{For example, the simplest TC model
has been ruled out many times.}

The remainder of this contribution is organized as follows. Sec. 2 contains a
description of the model as well as a discussion of the constraints that have
been imposed. In sec. 3 the resulting (s)particle spectrum will be discussed
in some detail; this section contains the only material that cannot be found
in our publication \cite{16}. In sec. 4 signals for sparticle production at
the LHC are treated for a few characteristic spectra; the emphasis will be
on methods to distinguish these spectra from each other as well as from the
kind of spectrum that has been studied previously. Finally, sec. 5 is devoted
to summary and conclusions.

\section*{2) The model}
\setcounter{footnote}{0}
As explained in the Introduction, I will assume a simple form for the
sparticle and Higgs spectrum at very high energy scales. In fact, one
ultimately hopes to describe all of SUSY breaking by a single parameter, the
equivalent of the electroweak symmetry breaking scale $(\sqrt{2} G_F)^{-1/2} =
246$ GeV. At present we are still far from this ambitious goal, so it is
prudent to allow at least a few free parameters to describe the spectrum. Here
I will follow the standard assumptions \cite{12} and introduce four
independent SUSY breaking parameters: $m_0^2$, which contributes to the
squared masses of all scalar bosons; a common gaugino mass \mhalf; and common
nonsupersymmetric trilinear and bilinear scalar interaction strengths $A$ and
$B$, respectively.\footnote{From supergravity or superstring theories one
might expect such a spectrum to emerge from an effective theory at scales
below the Planck scale ($\mpl = 2.4 \cdot 10^{18}$ GeV) or perhaps the string
compactification scale ($M_c \simeq 5 \cdot 10^{17}$ GeV). Here I will assume
that this ansatz is still valid at the GUT scale $\mx \simeq 2 \cdot 10^{16}$
GeV; in the scenario presented here the running of parameters from \mpl\ or
$M_c$ down to \mx\ is not expected to play a major role.} In addition one has
to introduce a supersymmetric Higgs(ino) mass $\mu$ in order to avoid the
existence of a (visible) axion and to give masses to both up-- and down--type
quarks. Altogether the masses of the two $SU(2)$ doublets of Higgs bosons
needed in any realistic SUSY model are then
\be \label{e1}
m_H^2(\mx) = m^2_{\bar H}(\mx) = m_0^2 + \mu^2(\mx).
\ee

One can show quite easily that spontaneous \sym\ breaking is not possible
as long as $m^2_H = m^2_{\bar H}$. Fortunately this degeneracy is lifted by
radiative corrections. The reason is that $\bar{H}$ only has Yukawa couplings
to up--type quarks, while $H$ only couples to down--type quarks and leptons.
Of course, the $t$ quark belongs to the former category, and has by far
the largest Yukawa coupling of all SM fermions (unless $\tanb \gg 1$; see
below). This implies that radiative corrections from Yukawa interactions will
be much larger for $m^2_{\bar H}$ than for $m_H^2$. The crucial point is that
these corrections {\em reduce} the (running) squared mass when going to lower
energy scales, eventually leading to nonzero vevs for $H$ and $\bar{H}$. As
already emphasized in the Introduction, this mechanism will only work if
$\log \mx / \mz \gg 1$ and the top Yukawa coupling $h_t$ is
large.\footnote{Strictly speaking, any nonzero $h_t$ will give a finite region
of input parameter space where radiative gauge symmetry breaking can be
achieved. However, this region would have been small, i.e. some fine--tuning
would have been needed, if $h_t$ were much smaller than the gauge couplings.
Radiative symmetry breaking is therefore more natural for heavy top.}

The running of the soft SUSY breaking parameters as well the of the gauge and
Yukawa couplings is described by a set of renormalization group equations
(RGE) \cite{17}. As a consequence of the assumption of minimal particle
content the model contains no intermediate scales between \mx\ and the weak
scale. The RGE therefore allow to uniquely determine the values of parameters
at the weak scale from the input parameters at \mx. Of course, at the weak
scale certain equalities have to be satisfied. First of all, we know that
\be \label{e2}
\frac {g^2 + g'^2} {2} \left( \langle H^0 \rangle^2 + \langle \bar{H}^0
\rangle^2 \right) = M_Z^2.
\ee
It is often convenient to introduce the weak scale parameter $\tanb \equiv
\langle \bar{H}^0 \rangle / \langle H^0 \rangle$. With the help of this
parameter, eq.(\ref{e2}) can be solved for $\mu^2$ at the weak scale
$Q_0$:\footnote{Eq.(\ref{e3}) is valid if the tree--level Higgs potential is
used with running parameters to include radiative correctios $\propto \log
(\mx/Q_0)$. However, it is always possible to chose the scale $Q_0$
where the RG running is terminated such that corrections to eq.(\ref{e3})
are small even when the full 1--loop Higgs potential is used \cite{18,18a},
which is almost independent of $Q_0$. The same choice of $Q_0$ also minimizes
corrections to the mass of the pseudoscalar Higgs boson.}
\be \label{e3}
\mu^2 (Q_0) = \frac {m_1^2(Q_0) - m_2^2(Q_0)} {\tan^2 \beta - 1} - \frac{1}{2}
M_Z^2,
\ee
where $m_1^2, m_2^2$ are the nonsupersymmetric contributions to the squared
Higgs masses, i.e. $m_H^2 = m_1^2 + \mu^2, \ m^2_{\bar H} = m_2^2 + \mu^2$.
For heavy top, $m_2^2(Q_0)$ is negative, giving a positive (and usually quite
large) contribution to $\mu^2$. A similar equation determines $B \cdot \mu$ at
scale $Q_0$ in terms of $m_1^2, m_2^2$ and \tanb.

Having fixed \mz, we are left with the parameters $m_0, \mhalf$ and $A$ (at
the GUT scale) as well as \tanb\ (at the weak scale). In addition the mass
$m_t$ of the top quark is an important parameter, since the top Yukawa
coupling plays a vital role in radiative gauge symmetry breaking. At this
point the assumption of a minimal $SU(5)$ GUT helps to further reduce the
number of free parameters. The reason is that minimal $SU(5)$ implies the
equality of $b$ and $\tau$ Yukawa couplings at scale \mx. As has been shown
by several groups \cite{19}, this can only be brought into agreement with the
experimentally measured ratio $m_b / m_{\tau}$ if either $h_t$ is close to its
upper bound or if $h_b \simeq h_t$, which implies $\tanb \simeq
m_t/m_b$.\footnote{This conclusion remains valid if $h_b/h_{\tau}$ is allowed
to vary by $\sim 10 \%$ from unity due to threshold effects or ``Planck
slop".} The second choice not only necessitates some finetuning in this
minimal scenario \cite{20}, it also makes it more difficult to satisfy proton
decay constraints (see below). I therefore only consider the first solution
here. Within the precision of a 1--loop calculation it can simply be
implemented by taking
\be \label{e4}
h_t (\mx) = 2.
\ee
This ensures that at low energies $h_t$ is very close to its infrared
quasi fixed point \cite{19}, which implies
\be \label{e5}
m_t(m_t) \simeq 190 \ {\rm GeV} \cdot \sin \! \beta,
\ee
where $m_t(m_t)$ is the running ($\overline{\rm MS}$) top mass at scale $m_t$;
the on--shell (physical) top mass is about 5\% larger. Eq.(\ref{e5}) reduces
the number of free parameters from five to four in this scenario. Assuming
$m_t$(pole)$\leq 185$ GeV, as indicated by LEP data \cite{20a} if the Higgs is
light, then implies
\be \label{e6}
\tanb \leq 2.5.
\ee

In addition to the relations discussed so far a number of conditions has to
be satisfied. These can all be expressed in terms of inequalities, i.e. they
define allowed ranges of parameter values rather than determining them
uniquely. One important constraint emerges from the requirement that nucleons
should be sufficiently long--lived. In minimal SUSY $SU(5)$ the main
contribution to nucleon decay comes from the exchange of the fermionic
superpartners of the $SU(5)$ partners of the elw. Higgs bosons, i.e. from
higgsino triplet exchange \cite{21}. The reason is that the corresponding
diagrams are only suppressed by one power of a mass ${\cal O}(\mx)$, as
compared to two such powers for $SU(5)$ gauge boson exchange. However, while
higgsino exchange suffices to violate both baryon and lepton number, it leads
to two sfermions, rather than two fermions, in the final state. These
sfermions have to be transformed into a lepton and an anti--quark by gaugino
(mostly chargino) exchange, i.e. the decay only occurs at 1--loop level. The
matrix element therefore contains a so--called dressing loop function, which
scales like $\mhalf/m_0^2$ for $m_0 \geq \mhalf$. Altogether one thus has
\cite{21}
\be \label{e7}
{\cal M} ( p \rightarrow K \nu) \propto \frac {\tan \! \beta} {m_{\tilde{H}_3}}
\frac {m_{1/2}} {m_0^2},
\ee
where $m_{\tilde{H}_3}$ is the mass of the Higgsino triplet, and the factor
\tanb\ appears since Yukawa couplings to $d$ and $s$ quarks grow $\propto
\tanb$. The experimental lower bound on the proton lifetime $\tau_p$ gives
an upper bound on the matrix element (\ref{e7}); assuming conservatively that
$m_{\tilde{H}_3}$ could be as much as ten times larger than the scale \mx\
where the gauge couplings meet, and taking into account that we are only
interested in rather small values of \tanb, eq.(\ref{e6}), a conservative
interpretation of the constraint imposed by the bound on $\tau_p$ is
\cite{21}\footnote{Strictly speaking the Higgs sector of minimal $SU(5)$ is
not realistic, since it leads to wrong predictions for the masses of SM
fermions of the first two generations. Adding new Higgs fields and/or
non--renormalizable interactions to solve this problem can in general also
change predictions for $\tau_p$. However, refs.\cite{21} already use the
physical quark masses as input. Furthermore, the constraint (\ref{e8}) is
really rather lenient; it can therefore be expected to hold even in slightly
extended models.}
\be \label{e8}
m_0 \geq \min(300 \ {\rm GeV}, \ 3 \mhalf).
\ee

Another important constraint can be derived from the requirement that relic
LSPs produced in the Big Bang do not overclose the universe, in which case it
would never have reached its present age of at least $10^{10}$ years. In
minimal SUGRA the lightest supersymmetric particle (LSP) is always the
lightest of the four neutralino states. Moreover, the large value of $h_t$,
eq.(\ref{e4}), and small \tanb\ (\ref{e6}) imply that $|\mu(Q_0)|$ is quite
large, see eq.(\ref{e3}). The LSP is therefore always a gaugino (mostly bino)
with small higgsino component. The LSP relic density is essentially inversely
proportional to its annihilation cross section, summed over all accessible
channels. Gaugino--like LSPs mostly annihilate into $f \bar{f}$ final states
\cite{22}, where $f$ stands for any SM fermion with mass below that of the
LSP. This final state is accessible via $\tilde{f}$ exchange in the $t$ or
$u$ channel, as well as via the exchange of the $Z$ boson or one of the
neutral Higgs bosons in the $s$ channel. However, the constraint (\ref{e8})
implies that the $\tilde{f}$ exchange contribution is strongly suppressed, due
to the large sfermion masses. (Recall that all sfermions get a contribution
$+m_0^2$ to their squared masses at scale \mx.) Moreover, since both $m_0$
and $|\mu|$ are large, most Higgs bosons are very heavy. Finally, the
LSP--LSP--$Z$ coupling needs two factors of the small higgsino component of
the LSP, while the LSP--LSP--Higgs couplings only need one such factor.
Therefore the only potentially large contribution to LSP annihilation comes
from the exchange of the light neutral Higgs boson $h^0$. Fortunately it has
recently been shown \cite{23} that this contribution suffices to reduce the
LSP relic density to acceptable values for a substantial range of LSP masses,
{\em provided} it is below $m_h/2$. The upper bound (\ref{e6}) on \tanb\
implies that $m_h \leq 110$ GeV even after radiative corrections \cite{24,18a}
are included, while Higgs searches at LEP imply \cite{25} $m_h \geq 63$ GeV.
(The couplings of $h^0$ are very similar to that of the SM Higgs boson in the
given scenario.) Since the mass of a bino--like neutralino is about $0.4
\mhalf$, this implies
\be \label{e9}
60 \ {\rm GeV} \leq \mhalf \leq 130 \ {\rm GeV}.
\ee

It should be emphasized that such a strong upper bound on \mhalf\ only holds
in this specific scenario. If we give up on the unification of $b$ and $\tau$
Yukawa couplings, we can allow smaller $h_t$ and/or larger \tanb, leading to
smaller values of $|\mu(Q_0)|$, and hence a larger higgsino component of the
LSP and stronger annihilation into final states containing Higgs and/or gauge
bosons. If we allow for a sufficiently non--minimal GUT Higgs sector the bound
on the proton lifetime can be satisfied even if the constraint (\ref{e8}) is
violated, allowing for efficient LSP annihilation through $\tilde{f}$
exchange. Finally, if we allow new, $R-$parity violating interactions, relic
LSPs could have decayed a long time ago, and no constraint on LSP
annihilation could be given.

Eqs.(\ref{e4}), (\ref{e6}), (\ref{e8}) and (\ref{e9}) describe the basic
allowed parameter space of the model, except for the $A$ parameter. The
bounds on this parameter are intimately linked to to the details of the
(s)particle spectrum, which is the topic of the next section.

\setcounter{footnote}{0}
\section*{3) The spectrum}
In this section I discuss the sparticle and Higgs spectrum of the model
defined in the previous section, with emphasis on features that are relevant
for ``new physics'' searches at colliders; see also refs.\cite{26,27} for
recent discussions of the spectrum in mSUGRA models with top Yukawa coupling
close to its fixed point.

A general feature of all mSUGRA models is that the Higgs(ino) mass parameter
$\mu$ is {\em not} an independent variable, but can be computed from the
SUSY breaking parameters, $m_t$ and \tanb\ (as well as \mz, of course), as
described by eq.(\ref{e3}). In general, the r.h.s. of that equation is a
complicated function of all input parameters, which has to be computed by
solving the relevant RGE \cite{17} numerically. However, the following
analytical expression are often sufficient for practical purposes:
\ben \label{e10} \beq
m_1^2(Q_0) &\simeq m_0^2 + 0.5 m_{1/2}^2; \label{e10a} \\
m_2^2(Q_0) &\simeq m_1^2(Q_0) - \frac{X_2}{\sin^2 \beta}, \ \ {\rm where}
\label{e10b} \\
X_2 &= \left( \frac {m_t} {150 \ {\rm GeV}} \right)^2 \left\{ 0.9 m_0^2 + 2.1
m^2_{1/2} \right. \nonumber \\ & \left. \hspace*{3cm} +
\left[ 1 - \brac^3 \right] \left( 0.24 A^2 + A \cdot \mhalf \right) \right\}
\label{e10c} \eeq \een
Eqs.(\ref{e10}) reproduce the exact numerical results to 10\% or better if
$Q_0$ is around 350 GeV (which corresponds \cite{18a} to squark masses around
600 GeV), and if $h_b \ll 1$, which is always true here due to the upper bound
(\ref{e6}) on \tanb. Note that $m_t$ in eqs.(\ref{e10}) is the running top
mass $m_t(m_t)$. The expression in square brackets in eq.(\ref{e10c}) will
therefore always be small if $h_t$ is close to its fixed point, see
eq.(\ref{e5}). The resulting very weak dependence of $X_2$ on $A$ has also
been observed in ref.\cite{26}.\footnote{Eqs.(\ref{e10}) are valid also away
from the IR fixed point of $h_t$. They differ from the analytical expressions
given in ref.\cite{26} partly because in that paper $Q_0 = \mz$ has been
assumed, which is not appropriate \cite{18,18a} for the heavy scalar spectrum
implied by the condition (\ref{e8}). (All numerical results in ref.\cite{26}
use the 1--loop corrected potential, and hence do not depend significantly on
the choice of $Q_0$.) Since most parameters run much more quickly just above
$Q_0$ than just below \mx\ changing $Q_0$ by a factor 4 is not completely
irrelevant. Notice also that ref.\cite{26} uses the opposite sign convention
for $A$.}

Note that eqs.(\ref{e10}) imply that $m_2^2(Q_0) < 0$ if $m_t / \sin \! \beta
> 158$ GeV; this is certainly true in the given scenario. Specifically, if
$h_t$ is close to the fixed point one has
\be \label{e11}
\mu^2(Q_0) > m_0^2 \frac {1 + 0.44 \tan^2 \beta} { \tan^2 \beta - 1}
> 0.72 m_0^2,
\ee
where I have used (\ref{e6}) in the second inequality; the lower bound
(\ref{e8}) on $m_0$ then implies that the LSP is indeed always a very pure
gaugino, as stated in the previous section. It also means that the two heavier
neutralinos and the heavier chargino, whose masses are all very close to
$|\mu(Q_0)|$, will be very difficult to detect: Their production cross section
at the LHC is much too small to yield a viable signal, while they are too
heavy to be produced at all at a linear \epem\ collider with $\sqrt{s} \simeq
500$ GeV, which is likely to be the next high--energy \epem\ collider.

Another useful identity is \cite{18a}
\be \label{e12}
m_P^2 = \frac{ m^2_{\tilde{\nu}} + \mu^2(Q_0) }{\sin^2 \beta},
\ee
where
\be \label{e13}
m^2_{\tilde{\nu}} \simeq m_0^2 + 0.5 m_{1/2}^2 + 0.5 M_Z^2 \cos \! 2 \beta
\ee
is the squared sneutrino mass and $m_P$ is the mass of the pseudoscalar Higgs
boson, which is almost degenerate with the charged and heavy neutral scalar
Higgs bosons if $M_P^2 \gg M_Z^2$ \cite{27a}. Note that eq.(\ref{e12}) is {\em
exact}, unlike eqs.(\ref{e10}), up to corrections of order $h_b^2$. In our
fixed point scenario eq.(\ref{e12}) and the bounds (\ref{e8}) and (\ref{e11})
imply that $m_P$ can easily exceed 1 TeV even for quite modest values of
$m_0$; such heavy SUSY Higgses are very difficult to detect (or even produce)
experimentally. On the other hand, this also implies that the couplings of the
light neutral scalar Higgs boson $h^0$ to SM fermions and gauge bosons are
practically identical to those of the SM Higgs \cite{27a}, so that $h^0$ will
be produced copiously at the LHC, at future \epem\ linacs, and perhaps even at
the second stage of LEP.\footnote{While by construction the invisible decay of
$h^0$ into two LSPs is always allowed in this model, the corresponding
branching ratio is very small, being proportional to the square of the small
higgsino component of the LSP.}

Since we have $m_0^2 \gg m^2_{1/2}$ in our model, all sleptons have very
similar masses; eqs.(\ref{e13}) and (\ref{e8}) then imply that they are most
likely too heavy to be detectable at the LHC \cite{28}, while they cannot be
produced at all at a 500 GeV \epem\ collider.

Clearly the best chance for a decisive test of the model is given by the
upper bound (\ref{e9}) on \mhalf. Since $|\mu(Q_0)|$ is so large the masses of
the lighter chargino and the next--to--lightest neutralino lie within a few
GeV of the low--energy $SU(2)$ gaugino mass $M_2 \simeq 0.82 \mhalf$. In
particular, the lighter chargino always lies below 105 GeV \cite{15}, offering
a good chance for its discovery at LEP, especially if the machine energy can
be boosted beyond the currently foreseen value of 176 GeV.

The situation is slightly more complicated for the gluino. The bound
(\ref{e9}) implies a rather low upper bound for the running ($\overline{MS}$
or $\overline{DR}$) gluino mass, $\mgl(\mgl) \simeq 2.5 \mhalf \leq 330$ GeV.
However, it has recently been pointed out \cite{29} that the on--shell gluino
mass can be substantially larger than this, especially if squarks are heavy:
\be \label{e14}
\mgl ({\rm pole}) \simeq \mgl(\mgl) \left[ 1 + \frac {\alpha_s(m_{\tilde g})}
{\pi} \left( 3 + \frac{1}{4} \sum_{\tilde q} \log \frac {m_{\tilde q}}
{m_{\tilde g}} \right) \right],
\ee
where I have only included the leading logarithmic correction from
squark--quark loops. Requiring somewhat arbitrarily
\be \label{e15}
m_0 \leq 1 \ {\rm TeV}
\ee
in order to avoid excessive finetuning in eq.(\ref{e3}), we see that the
``threshold correction" (\ref{e14}) can change the gluino mass by about 30\%
if \mhalf\ is close to its lower bound and $m_0$ is at its upper bound. This
can change both the cross section and the signal for gluino pair production
quite significantly. Including the correction (\ref{e14}), the bound
(\ref{e8}) implies $\mgl($pole$) \leq 400$ GeV if (\ref{e15}) is imposed.

The lower bound (\ref{e8}) on $m_0$ also implies that the squarks of the
first two generations are significantly heavier than the gluinos; they will
thus dominantly decay into $\tilde{g} + q$, although the left--handed ($SU(2)$
doublet) squarks also have ${\cal O}(10\%)$ branching ratios into an elw.
gaugino and a quark. The masses of the first two generations of squarks lie
between about 330 GeV and a little above 1 TeV in this model, where the upper
bound merely reflects the somewhat arbitrary constraint (\ref{e15}).

So far the spectrum resembles a special case of the type of models whose
signatures were discussed in the existing literature \cite{1,11}, with heavy
squarks and sleptons, large $|\mu|$ and $m_P$, but rather light gluino and
elw. gauginos. This is not surprising, since in these ealiers studies $\mu$
and $m_P$ were considered to be free parameters; they could thus be chosen to
be large, whereas the present model {\em requires} them to be large. However,
as already remarked in the Introduction, in these papers all squarks were
also assumed to have the same mass at the weak scale. It is here that the
present model makes specific predictions which cannot be mimicked by these
earlier treatments, in spite of the larger number of free parameters.

The reason is that the same kind of radiative corrections that reduce the
Higgs mass parameter $m_2^2$ to negative values also reduce the masses of the
stop and left--handed sbottom squarks, as compared to the masses of first
generation squarks. Specifically, one finds
\ben \label{e16} \beq
m^2_{\tilde{b}_L} &= m^2_{\tilde{t}_L} \simeq m^2_{\tilde q} - \frac {X_2}
{3 \sin^2 \beta} ; \label{e16a} \\
m^2_{\tilde{t}_R} &\simeq m_{\tilde q}^2 - \frac {2 X_2} {3 \sin^2 \beta},
\label{e16b} \eeq \een
where
\be \label{e17}
m^2_{\tilde q} \simeq m_0^2 + 6 m^2_{1/2}
\ee
is a typical first generation squark mass (at scale $Q=m_{\tilde q}$), and
$X_2$ has been given in eq.(\ref{e10c}).

Since for the small values of \tanb\ of interest here $\tilde{b}_L -
\tilde{b}_R$ mixing is almost negligible, $\tilde{b}_L$ is to good
approximation the mass eigenstate $\tilde{b}_1$, with mass given by
eq.(\ref{e16a}). The ratio $m_{\tilde{b}_1} / m_{\tilde{u}_L}$ (including
small mixing effects and $D-$terms) is shown in fig. 1 as a function of
$A_0 \equiv A/m_0$, for various values of $m_0$ and \mhalf\ and $\mu <
0$.\footnote{Note that eq.(\ref{e5}) only determines the absolute value of
$\mu$. In this paper I assume $\mhalf > 0, \ \tanb > 0$ without loss of
generality; however, $A, \ B$ and $\mu$ must then in general all be allowed to
have either sign. The sign conventions for $\mu$ and \mhalf\ used here are the
same as in refs.\cite{27a,30}.} We observe that the dependence on $A$ is
indeed quite weak here, as claimed earlier. The allowed range for $A_0$ is
determined by various constraints: The squared mass of the lightest stop
eigenstate (see below) must be larger than $+ (45 \ {\rm GeV})^2$; the scalar
potential at the weak scale should not have minima \cite{31} in the directions
$\langle \tilde{\tau}_R \rangle = \langle \tilde{\tau}_L \rangle
= \langle H^0 \rangle$ or $\langle \tilde{t}_R \rangle = \langle
\tilde{t}_L \rangle =  \langle \bar{H}^0 \rangle $ that are deeper than the
desired minimum where only $\langle H^0 \rangle$ and $\langle \bar{H}^0
\rangle$ are nonzero; and the potential must be bounded from below at the GUT
scale, which requires $m_0^2 + \mu^2(\mx) \geq 2 | B(\mx) \mu(\mx)|$.


The ratio $m_{\tilde{b}_1}/m_{\tilde{u}_L}$ does depend somewhat on the ratio
$\mhalf/m_0$, however, being maximal where $\mhalf/m_0$ is minimal (and vice
versa). We see that in our fixed--point scenario $m_{\tilde{b}_1}$ is reduced
by typically 20 to 30\% compared to first generation squark masses. This can
be quite significant, since partial widths for three--body decays of gluinos
or elw. gauginos that involve squark exchange scale approximately like the
inverse fourth power of the squark mass. A reduction of the squark mass by
20\% (30\%) therefore leads to an increase of the corresponding partial width
or branching ratio by a factor of 2.1 (2.9). This leads to $b-$rich final
states, as will be discussed in more detail in sec. 4.

Eq.(\ref{e16b}) implies that the mass of the $\tilde{t}_R$ current state is
reduced even more than $m_{\tilde{b}_L}$. Moreover, $\tilde{t}_L -
\tilde{t}_R$ mixing is not negligible; it reduces the mass \mst\ of the
lighter $\tilde{t}$ mass eigenstate even further. In the convention of
ref.\cite{30} the $\tilde{t}$ mass matrix is given by \cite{32}:
\be \label{e18}
{\cal M}^2_{\tilde t} = \mbox{$ \left( \begin{array}{cc}
m^2_{\tilde{t}_L} + m_t^2 + 0.35 M_Z^2 \cos \! 2 \beta
& - m_t ( A_t + \mu \cot \! \beta ) \\
- m_t ( A_t + \mu \cot \! \beta ) &
m^2_{\tilde{t}_R} + m_t^2 + 0.16 M_Z^2 \cos \! 2 \beta
\end{array} \right) $}.
\ee
At scale \mx, $A_t = A$, but it is subject to radiative corrections due to
both gauge and Yukawa interactions. For small \tanb, one has approximately:
\be \label{e19}
A_t(Q_0) \simeq A \left[ 1 - \brac^2 \right] 
+ \mhalf \left[ 3.5 - 1.9 \brac^2 \right],
\ee
where I have again assumed $Q_0 \simeq 350$ GeV. Right at the fixed point of
$h_t$, where eq.(\ref{e5}) becomes exact, the weak--scale value of $A_t$ is
again independent of $A$ \cite{26}. However, even for values of $h_t(\mx)$ as
large as 2, \mst\ can still show substantial $A-$dependence. The reason is
that in the expression for \mst\ strong cancellations occur between the
contributions from diagonal and off--diagonal entries of the stop mass matrix;
relatively small changes of these elements, of the size shown in fig. 1, can
therefore have sizable effects on \mst. This is especially true for $\mu > 0$,
since there all effects go in the same direction: Increasing $A$ increases
$X_2$, which simultaneously reduces $m^2_{\tilde{t}_L}$ and
$m^2_{\tilde{t}_R}$ in the diagonal entries, eq.(\ref{e16}), and increases the
off--diagonal elements by increasing $\mu$, see eqs.(\ref{e3}) and
(\ref{e10}); note that $A_t$ and $\mu$ have the same sign in this case.

\mst\ also depends quite strongly on $m_t$ in the fixed--point scenario. The
reason is that a larger $m_t$ implies larger \tanb, see eq.(\ref{e5}), and
hence smaller $|\mu|$, eq.(\ref{e3}); in addition, the off--diagonal term
in the stop mass matrix (\ref{e18}) is reduced, due to the explicit $\cot \!
\beta$ factor. Therefore larger $m_t$ also imply larger \mst here. (This is
not generally true if $h_t$ is significantly below its fixed point \cite{30}.)

In all of parameter space allowed in our model one finds that \mst\ is close
to or even well below the gluino mass. This is partly due to the correction
(\ref{e14}).\footnote{The difference between running and on--shell masses is
much smaller \cite{33} for squarks than for gluinos, since squarks are color
triplets (rather than octets) and no sum over flavours occurs, unlike in
eq.(\ref{e14}).} Not surprisingly, this statement remains true \cite{26,27}
if the constraint (\ref{e8}) from the proton lifetime is not imposed, i.e.
if larger ratios $\mhalf/m_0$ are allowed, as long as $h_t$ is close to its
fixed point. Light stops are hence a quite generic prediction of SUGRA models
with large $h_t$. Their phenomenology depends on whether or not they can be
produced in gluino decays (i.e., whether $\mgl > m_t + \mst$), and also
whether the decay $\tilde{t}_1 \rightarrow b + \tw_1$ is allowed, where
$\tw_1$ is the lighter chargino; if (and only if) this decay is
forbidden, $\tilde{t}_1 \rightarrow c + \tilde{Z}_1$ via loop diagrams
\cite{34}.


The ordering of \mgl, \mst, $m_t$ and $m_{\widetilde{W}_1}$ depends quite
strongly on $A$, \mhalf\ and $m_t$, as well as on the sign of $\mu$. This is
demonstrated by figs. 2 a--d, which show regions in the $(A_0,\mhalf)$ plane
for $m_0=500$ GeV; results for other values of $m_0$ allowed by (\ref{e8}) are
quite similar. Figs. 2a,b are for $m_t(m_t) = 160$ GeV, corresponding to
$m_t($pole$)=168$ GeV, while c,d are for $m_t(m_t)=175$ GeV or $m_t($pole$) =
183$ GeV; $\mu$ has been chosen positive in a,c and negative in b,d. In all
these figures the regions outside the dotted lines are excluded by the
constraints described in the discussion of fig. 1. Along the solid line one
has $\mgl = \mst + m_t$, i.e. on one side of this line $\tilde{g} \rightarrow
\st + t$ decays are allowed; since all other squarks are considerably heavier,
this is the only possible two--body decay mode of the gluino and will hence
have a branching ratio of nearly 100\% if allowed at all. Similarly, the long
dashed lines separate regions where $\tilde{t}_1 \rightarrow \tw_1 + b$
is open, in which case it has a branching ratio very close to 100\%, from
those where $\tilde{t}_1 \rightarrow \tilde{Z}_1 + c$. The decay $\tilde{t}_1
\rightarrow t + \tilde{Z}_1$ is almost never allowed here (it opens up if both
\mhalf\ and $m_0$ are near their upper bounds); instead, the short dashed
lines are contours of constant $\mst = m_t$. In all cases \mst\ increases with
increasing \mhalf, and it generally also increases with decreasing $|A_0|$,
although the maximum (for given \mhalf) is not exactly at $A_0=0$.

In fig. 2a, $\tilde{g} \rightarrow \tilde{t}_1 + t$ decays are allowed
everywhere except within the small triangular region delineated by the solid
line. There are also substantial regions where $\tilde{t}_1 \rightarrow
\tw_1 + b$ is not possible, and $\mst < m_t$ almost everywhere except
for the small region enclosed by the the short dashed curve. Changing the
sign of $\mu$ (fig. 2b) or increasing $m_t(m_t)$ to 175 GeV (fig. 2c) changes
the situation quite drastically, however. The region where $\tilde{g}
\rightarrow \tilde{t}_1 + t$ is allowed is now confined to the narrow strips
between the solid and dotted lines, and $\mst < m_{\widetilde{W}_1} + m_b$ only
in the even narrower strips between the long dashed and dotted lines. In
contrast, the region where $\mst > m_t$ (above the short dashed lines) is now
quite large. Finally, if $m_t$ is close to its upper bound and $\mu < 0$,
fig. 2d, the gluino can never decay into $\tilde{t}_1 + t$, and $\tilde{t}_1$
always decays into $\tw_1+b$; moreover, now $\mst > m_t$ over most of the
allowd part of the $(A_0,\mhalf)$ plane.

Since $\tilde{t}_1$ and the gluino are by far the lightest strongly
interacting sparticles in our model, SUSY signals at the LHC (or other hadron
colliders) will obviously depend quite sensitively on which of the regions
depicted in figs. 2 is picked. This is the subject of the next section, where
these signals are discussed in more detail.

\section*{4) Signals at the LHC}
\setcounter{footnote}{0}
In this section I discuss the specific signatures produced by the kind of
spectrum described in the previous section. The signatures at \epem\ colliders
are rather straightforward: A light chargino (below 105 GeV), a rather light
neutral scalar Higgs boson (below 110 GeV), and a light \st\ (below $\sim 300$
GeV), all of which can be detected in a straightforward way.\footnote{The only
possible problem could occur if \st\ is very close in mass to the LSP, which
is in principle possible. The model also predicts the existence of a light
$\tilde{Z}_2$, with mass very close to $m_{\widetilde{W}_1}$, but its
production cross section at \epem\ colliders is small, since it is a rather
pure gaugino, so that $Z$ exchange is suppressed by small couplings, while
selectron exchange is suppressed by the large selectron mass.} Due to numerous
backgrounds the situation at hadron colliders is much more complicated.
Moreover, at these machines the largest signals come from the production of
strongly interacting sparticles. Since these are usually heavier than many
other sparticles they tend to decay via lengthy cascades \cite{11}. While this
makes signals more difficult to analyze, it also offers the opportunity to
determine various branching ratios, which greatly helps to distinguish between
different SUSY scenarios.

\begin{center}
{\bf Table 1:} Parameters and masses for six SUGRA cases A1, A2 and B1--B4. \\
\vspace*{0.2cm}
\begin{tabular}{|l||r|r|r|r|r|r|}
\hline
parameter & A1 & A2 & B1 & B2 & B3 & B4 \\
\hline
\hline
$m_0$          & 500 & 500 & 300 & 1000 & 400 & 1000 \\
$\mhalf $ & 120 & 130 & 60 & 70 & 130 & 130 \\
$A_0/m_0$      & 0.6 & 3.75 & 0.1 & 0.0 & 0.0 & 0.0 \\
$\tanb$     & 1.49 & 2.2 & 1.94 & 1.32 & 2.22 & 2.11 \\
$\mu$          & 697.1 & 580 & -313.3 & -1571.7 & 430 & 964.7 \\
$m_t$          & 160 & 175 & 170 & 155 & 175 & 175 \\
$m_{\tg}$      & 346 & 371 & 185 & 231 & 364 & 400 \\
$m_{\tq}$      & 568 & 578 & 328 & 1011 & 496 & 1039 \\
$m_{\tt_1}$    & 131 & 83.6 & 198 & 121 & 225 & 288 \\
$m_{\tt_2}$    & 521 & 500 & 304 & 781 & 470 & 799 \\
$m_{\tb_L}$    & 437 & 426 & 250 & 732 & 396 & 765 \\
$m_{\tl}$      & 508 & 501 & 305 & 1001 & 412 & 1005 \\
$m_{\tz_1}$    & 43.6 & 48.0 & 27.1 & 28.8 & 46.1 & 50.7 \\
$m_{\tz_2}$    & 85.1 & 93.5 & 63.5 & 59.8 & 89.3 & 99.9 \\
$m_{\tz_3}$    & 698 & 582 & 316 & 1572 & 432 & 966 \\
$m_{\tz_4}$    & 710 & 595 & 333 & 1577 & 450 & 973 \\
$m_{\tw_1}$    & 84.5 & 92.9 & 62.2 & 59.7 & 87.9 & 99.6 \\
$m_{\tw_2}$    & 707  & 593 & 331   & 1576 & 448 & 972 \\
$m_{h^0}$   & 85.2 & 99.7 & 61.6 & 86.0 & 91.3 & 101 \\
$m_{H^0}$      & 1038 & 841 & 492 & 2339 & 652 & 1543 \\
$m_{P}$      & 1038 & 842 & 487 & 2342 & 649 & 1542 \\
$m_{H^\pm}$    & 1040 & 845 & 492 & 2343 & 653 & 1544 \\
\hline
\end{tabular}
\end{center}
\vspace*{0.7cm}

Here I summarize the results of ref.\cite{16}, where a full Monte Carlo study
of signals and backgrounds was performed, using the latest version of ISAJET
which contains the ISASUSY program package to compute sparticle masses and
decay branching ratios. The program includes initial and final state
showering, fragmentation, and crude detector modelling, where we took present
designs of LHC detectors as guidelines; see ref.\cite{35} for more details on
this program package.

A simulation of this type consumes a substantial amount of CPU
time.\footnote{A typical gluino pair event has ${\cal O}(1000)$ hadrons and
photons in the final state.} We therefore limited ourselves to  the study of
six spectra that can occur within SUGRA $SU(5)$, see table 1. We saw already
in the previous section that over a substantial region of parameter space
the gluino can decay into $t+\st$. The two `A' spectra were picked from
that region, with A1 being an example where $\st \rightarrow \tw_1+b$
while in case A2 the light stop decays into charm plus LSP. In the remaining
four `B' cases the gluino has no two--body decay modes; here we picked
points where $m_0$ and \mhalf\ are minimal or maximal: $(m_0, \ \mhalf) = $
(0.3 TeV, 60 GeV) (B1); (1 TeV, 70 GeV) (B2); (0.4 TeV, 130 GeV) (B3); and
(1 TeV, 130 GeV) (B4). These six spectra show all the features discussed
earlier: Light $\tilde{Z}_1, \ \tilde{Z}_2, \ \tw_1, \ h^0$ and
$\tilde{g}$; quite heavy squarks and sleptons, except for \st, with
$\tilde{b}_1$ significantly below first generation squarks; heavy
$\tilde{Z}_3, \ \tilde{Z}_4$ and $\tw_2$; and very heavy Higgs bosons
$P, \ H^0$ and $H^{\pm}$.

In order to see how mSUGRA signals differ from the signals of ``conventional
SUSY" models of the type considered in ref.\cite{11}, we also studied four
cases (labelled BTW1 through BTW4) where all squarks and sleptons are assumed
to be degenerate. In all `BTW' cases we took $\mgl=300$ GeV, and picked
$(m_{\tilde q},\ \mu)=$ (320 GeV, --150 GeV) (BTW1); (600 GeV, --150 GeV)
(BTW2); (320 GeV, --500 GeV) (BTW3); and (600 GeV, --500 GeV)
(BTW4).\footnote{Recall that $\mu$ is assumed to be a free parameter in this
``conventional" treatment. The {\em calculated} values of $\mu$ in the six
mSUGRA cases are listed in table 1.}

In table 2 SUSY event fractions as well as important sparticle decay branching
ratios are listed. In the `A' cases with very light \st, more than half of all
SUSY events are $\st \tilde{t}_1^*$ pairs, while in the four `B' cases, gluino
pairs constitute the most copiously produced supersymmetric final state. Pairs
of electroweak gauginos are produced only in a few percent or even few
permille of all SUSY events; their detection therefore necessitates a
dedicated search \cite{36}, as opposed to the ``generic SUSY search" presented
here.

Many of the branching ratios shown in table 2 can be understood directly from
the sparticle masses listed in table 1, keeping in mind that two--body final
states will always overwhelm three--body final states if both are accessible
at tree level. A few features are worth pointing out, however. For example, in
cases B1, B2 the $\lsp b \tilde{b}_L$ coupling is ``accidentally" suppressed.
The branching ratio for $\tilde{g} \rightarrow \lsp \bbbar$ is therefore
dominated by $\tilde{b}_R$ exchange, whose mass is not reduced compared to
first generation squark masses; therefore $\tilde{g} \rightarrow \lsp \bbbar$
is not enhanced significantly over $\tilde{g} \rightarrow \lsp d \bar{d}$ in
these cases. Nevertheless $\tilde{Z}_2 \rightarrow \lsp \bbbar$ is enhanced
considerably over $\tilde{Z}_2 d \bar{d}$. The reason is that virtual $h^0$
exchange diagrams, which contribute much more strongly to \bbbar\ final
states, are not negligible here. This can be understood from the observation
that $h^0$ exchange is only suppressed \cite{27a} by one power of the small
higgsino component of $\tilde{Z}_2$ or \lsp, while $Z$ exchange needs \cite{1}
two such powers, and $\tilde{f}$ exchange is suppressed by the large sfermion
masses.

$h^0$ exchange contributes negligibly to $\tilde{Z}_2 \rightarrow \lsp \eplem$
decays, so it reduces the corresponding branching ratio.\footnote{Note that,
at least in the limit $m_b \ll m_{\tilde{Z}_2}$, $h^0$ exchange does not
interfere with sfermion or $Z$ exchange; it therefore always contributes
positively.} The leptonic branching ratio of $\tilde{Z}_2$ can nevertheless
exceed considerably that of the $Z$ boson, even if (most) squarks and sleptons
have almost the same mass, as is the case here; this is largely due to
interference between $Z$ and sfermion exchange diagrams, which can however
also result in very small leptonic branching ratios for $\tilde{Z}_2$, see
cases B3 and B4. In contrast, the branching ratios of the light chargino very
closely track that of the $W$ bosons, since the $\tw_1 \lsp W$ couplings
get contributions from both the higgsino and $SU(2)$ gaugino components of
$\tw_1$ and \lsp, and are therefore much less suppressed than the
$\tilde{Z}_2 \lsp Z$ coupling. The only exception occurs in case A2, where the
two--body decay $\tw_1 \rightarrow \st + b$ is allowed and hence
completely dominates all $\tw_1$ decays.

\vspace*{0.7cm}
\noindent
{\bf Table 2:} (a) Fractions of SUSY particle pairs produced in $pp$ collisions
at the LHC; and (b) branching fractions of selected decay modes,
for six SUGRA cases A1, A2 and B1--B4, where $\tq$ stands for all squarks
except stops, and $\tx\tx$ stands for all possible
chargino and neutralino pairs. \\
\begin{center}
\begin{tabular}{|l|llllll|}
\hline
SUSY particles$\backslash$ Case & A1 & A2 & B1 & B2 & B3 & B4 \\
\hline
\hline
(a) Sparticle Pairs Produced & & & & & &\\
$\tg\tg$         & 0.30  & 0.093 & 0.72  & 0.74  & 0.44  & 0.73  \\
$\st\tilde{t}_1*$& 0.51  & 0.84  & 0.011 & 0.21  & 0.10  & 0.083 \\
$\tg\tq$         & 0.13  & 0.050 & 0.23  & 0.013 & 0.33  & 0.081 \\
$\tq\tq$         & 0.018 & 0.007 & 0.029 & 3$\times 10^{-4}$ & 0.067 & 0.005 \\
$\tw_1^\pm\tz_2$ & 0.018 & 0.006 & 0.004 & 0.019 & 0.027 & 0.066 \\
$\tx\tx$         & 0.026 & 0.009 & 0.007 & 0.027 & 0.042 & 0.088 \\
\hline
(b) Important Decay Modes & & & & & & \\
$\tg\to \tt_1 \bar{t},\bar{\tt_1}t$   & 1.0 & 1.0 & - & - & - & - \\
$\tg\to \tw_1^-t\bar{b},\tw_1^+b\bar{t}$ & 1.6$\times 10^{-4}$ &
           1.4$\times 10^{-4}$ & -     & 0.091 & 0.12 & 0.25 \\
$\tg\to \tw_1^-u\bar{d},\tw_1^+d\bar{u}$ & 4.6$\times 10^{-4}$ &
           2.9$\times 10^{-4}$ & 0.21  & 0.10 & 0.19 & 0.16 \\
$\tg\to \tz_1 d\bar{d}$   & 3.1$\times 10^{-5}$ & 1.9$\times 10^{-5}$
                         & 0.012 & 0.005 & 0.014 & 0.01 \\
$\tg\to \tz_1b\bar{b}$   & 6.8$\times 10^{-5}$ & 5.4$\times 10^{-5}$
                         & 0.012 & 0.006 & 0.035 & 0.018 \\
$\tg\to \tz_2b\bar{b}$   & 4.3$\times 10^{-4}$ & 3.9$\times 10^{-4}$
                         & 0.21  & 0.10  & 0.18  & 0.15 \\
$\tt_1\to \tw_1^+ b$     & 1.0 & -   & 0.95 & 1.0 & 0.93 & 0.29 \\
$\tt_1\to \tz_1 t$       & -   & -   & 0.05 & -   & 0.07 & 0.64 \\
$\tt_1\to \tz_2 t$       & -   & -   & -    & -   & -    & 0.07 \\
$\tt_1\to \tz_1 c$       & -   & 1.0 & -    & -   & -    & -    \\
$\tz_2\to\tz_1 d\bar{d}$ & 0.11  & 0.18   & 0.024
                         & 0.028  & 0.21   & 0.17 \\
$\tz_2\to\tz_1 b\bar{b}$ & 0.37  & 0.39   & 0.057
                         & 0.22  & 0.38   & 0.42 \\
$\tz_2\to\tz_1 e^+e^-$   & 0.047 & 0.018  & 0.14
                         & 0.12  & 0.007  & 0.012 \\
$\tw_1^+\to\tz_1 e^+\nu_e $ & 0.11 & 6.9$\times 10^{-5}$
                            & 0.11 & 0.11 & 0.11 & 0.11 \\
\hline
\end{tabular}
\end{center}
\vspace*{0.7cm}

It is by now quite well known that, in addition to the ``classical" missing
$E_T + $jets signature, sparticle production at hadron colliders can also give
rise to final states containing hard leptons \cite{11,37}. In the present
study a hard lepton is defined as an electron or muon with $p_T > 20$ GeV,
pseudorapidity $|\eta| < 2.5$, and visible (hadronic or electromagnetic)
activity in a cone $\Delta R = 0.3$ around the lepton less than 5 GeV.
Hadronic clusters with $E_T > 50$ GeV in a cone $\Delta R = 0.7$ are labelled
as jets. We can then define various mutually exclusive event classes:
\begin{itemize}
\item ``Missing $E_T$" events have no hard leptons, $n_j \geq 4$ jets, missing
$E_T > 150$ GeV, transverse sphericity $S_T > 0.2$, and total scalar
(calorimetric) $E_T > 700$ GeV. This last cut is not strictly necessary, but
greatly enhances signal/background.
\item ``$n$ lepton" events have exactly $n \ (\geq 1)$ hard leptons, and
missing $E_T > 100$ GeV. If $n=1$, we in addition required the scalar $E_T$ to
exceed 700 GeV, and demanded that the transverse mass computed from the
missing $p_T$ and the leptonic $p_T$ does {\em not} fall in the interval
between 60 and 100 GeV, where the background from real $W$ decay has a
Jacobian peak. For $n=2$ we distinguish opposite sign (OS) and same sign (SS)
events, and required total $E_T > 700$ GeV in the OS sample in order to
suppress $t \bar{t}$ backgrounds.
\end{itemize}

Cross sections after cuts for these final states are listed in table 3, for
the six mSUGRA cases, 4 `BTW' cases and leading sources of background, i.e.
$t \bar{t}$, $W+$jets and $Z+$jets events. (We checked that backgrounds from
$W^+W^-, \ c \bar{c}$ and \bbbar\ production are always very small.) The table
extends out to $n=4$ leptons, but the results for $n=4$ already suffer from
substantial statistical errors (we generated 50,000 events for each SUSY
spectrum, and several hundred thousand background events). We observe that the
missing $E_T$, SS and (except in case A2) $3l$ signals are all well above
background; the $1l$ and OS signals are also always larger than, or at least
similar to, backgrounds.\footnote{A warning: All cross sections in table 3
are not only subject to unknown NLO QCD corrections, but also depend
sensitively on the resolution and coverage of the calorimeters, since both
signal and background are backed up against the missing $E_T$ cut. Notice that
the predicted missing $E_T$ also depends on, for example, the amount of
initial state radiation that is being generated. One will therefore have to
understand both the detector and ``ordinary" QCD events in some detail before
quantitative studies of absolute rates can be undertaken with some confidence.
However, the {\em ratios} of various cross sections, including
signal/background, will hopefully be rather robust against future
refinements.}

The size of the missing $E_T$ cross section is mostly determined by the gluino
mass; direct $\st \tilde{t}_1^*$ production contributes little to the signal
after cuts even in case A2, mostly because we demand at least 4 jets here.
However, due to the potentially sizable contribution from $\tilde{g} \tilde{q}
$ production shown in table 2, the masses of first generation squarks also
play a role. Moreover, even this relatively robust signal can change by a
factor of more than 2 depending on gluino branching ratios: Even though case
A2 has a slightly heavier gluino than case A1, and hence an almost 50\%
smaller gluino pair cross section, it produces a two times stronger missing
$E_T$ signal than case A1 does. The reason is that $\st \rightarrow \lsp + c$
decays give much harder LSPs than $\st \rightarrow \tw_1 + b \rightarrow
\lsp + q \bar{q}' + b$ decays do. Similarly, the missing $E_T$ signal in case
B2 is almost as strong as in case B1, inspite of the more than three times
smaller sum of $\tilde{g} \tilde{g}$ and $\tilde{g} \tilde{q}$ cross sections.
This is mostly due to a large (22\%) branching ratio for $\tilde{g}
\rightarrow \lsp + g$ loop decays. ($\tilde{g} \rightarrow \tilde{Z}_2 + g$
decays also have a branching ratio of about 22\%, and about 20\% of all
$\tilde{Z}_2$ decay invisbly into $\lsp \nu \bar{\nu}$ in this case.) These
\lsp\ produced in $\tilde{g}$ two--body decays are very energetic, much more
so than the LSPs produced at the end of a cascade. The branching ratios for
these loop decays of the gluino are enhanced in case B2 because ordinary
squarks are very heavy, while \st\ (which contributes to loops) is quite
light.

\vspace*{0,7cm}
\noindent
{\bf Table 3:} Cross sections in $pb$ for various event topologies after cuts
described in the text, for $pp$ collisions at $\sqrt{s}=14$~TeV.
The various SUGRA cases are listed in the first column. The OS/SS ratio
is computed with the OS dilepton sample {\it before} the scalar $E_T$ cut.

\begin{center}
\begin{tabular}{|c|ccccccc|}
\hline
case & $\eslt$ & $1\ \ell$ & $OS$ & $SS$ & OS/SS & $3\ \ell$ & $4\ \ell$ \\
\hline
\hline
A1 & 24.6 & 36.2 & 5.4 & 3.7 & 2.0 & 1.2 & 0.017 \\
A2 & 48.0 & 31.4 & 1.5 & 2.1 & 1.2 & $<0.02$ & $<0.02$ \\
B1 & 79.1 & 76.8 & 11.9 & 3.4 & 6.9 & 1.7 & 0.17 \\
B2 & 67.7 & 37.5 & 9.0 & 2.4 & 7.7 & 0.8 & 0.1 \\
B3 & 51.8 & 21.2 & 1.4 & 0.6 & 3.3 & 0.09 & $<0.01$ \\
B4 & 20.1 & 10.1 & 1.1 & 0.4 & 3.5 & 0.1 & $<0.004$ \\
BTW1 & 105 & 39.8 & 2.8 & 1.4 & 3.1 & 0.21 & 0.03 \\
BTW2 & 57.3 & 22.5 & 2.2 & 0.85 & 3.6 & 0.14 & $<0.02$ \\
BTW3 & 96 & 58 & 10.9 & 2.9 & 6.7 & 1.5 & 0.06 \\
BTW4 & 52.3 & 23.9 & 2.3 & 0.9 & 3.7 & 0.2 & $<0.02$ \\
$t\bar t(160)$ & 2.9 & 8.0 & 1.1 & 0.01 & 640 & $<0.004$ & $<0.004$ \\
$W+jet$ & 0.6 & 3.8 & 0.29 & -- & & -- & -- \\
$Z+jet$ & 0.6 & 0.2 & 0.02 & -- & & -- & -- \\
$total\ BG$ & 4.3 & 12.02 & 1.411 & 0.01 & & $<0.004$ & $<0.004$ \\
\hline
\end{tabular}
\end{center}
\vspace*{0.7cm}

The size of the $1l$ signal is generally roughly comparable to that of the
missing $E_T$ signal, but backgrounds are about three times larger here. Real
top quarks are very efficient in producing hard leptons (and missing $E_T$,
needed to pass the cut $E\llap/_T > 100$ GeV); in case A1 the $1l$ signal
is therefore actually larger than the missing $E_T$ signal. Notice also that
the effect of increasing $m_0$ to 1 TeV is now about the same for cases B1/B2
and B3/B4, reducing the signal by approximately a factor of two. Many ``$1l$"
events in cases B1, B2 actually come from $\tilde{Z}_2 \rightarrow \lsp l^+
l^-$ decays where one of the leptons failed to pass the cuts. The large
leptonic branching ratio of $\tilde{Z}_2$ in these cases also leads to quite
large OS $2l$ signals, large OS/SS ratios, and sizable $3l$ signals. Case A1
also has substantial OS and $3l$ signals, but the OS/SS ratio is much smaller
here, since $\tilde{g} \tilde{g}$ events produce $tt$ and $\bar{t} \bar{t}$
final states with equal abundance as $t \bar{t}$ final states. Case A2 has
very few $n \geq 3$ lepton events, since basically all gluinos decay into $t +
\bar{c} + \lsp$ (or the charge conjugate thereof) here, giving at most one
hard lepton per gluino. Cases B3 and B4 give smaller leptonic signals, due to
the larger gluino mass; they differ from each other by a factor of two in the
missing $E_T$ and (marginal) $1l$ signals, but become more similar to each
other as more leptons are required, mostly due to the sizable $\tilde{g}
\rightarrow \tw_1 t b$ branching ratio in case B4.

Finally, the ``conventional" BTW cases can produce even larger missing $E_T$
signals than our mSUGRA cases, since we allowed squarks to lie just above
gluinos here, leading to large $\tilde{g} \tilde{q}$ production rates. Case
BTW3 also has a large leptonic branching ratio of $\tilde{Z}_2$ (12.5\% per
generation), and hence large cross sections for the $n$ lepton final states;
for the other BTW cases the leptonic branching ratio of $\tilde{Z}_2$ is
close to that of the $Z$ boson.

This discussion shows that the counting rates in the six or seven independent
signal channels listed in table 3 already go a long way towards distinguishing
the various mSUGRA cases from each other as well as from the `BTW' cases.
Some ambiguities remain, however; for example, cases B1 and BTW3 are are very
similar at this level (except perhaps in the $4l$ signal, but statistics is
poor here). The remaining ambiguity can be resolved by looking at the events
within one signal class in more detail. An example is shown in table 4, where
event fractions for the missing $E_T$ sample are shown, together with the
average total scalar $E_T$ and average missing $E_T$ per event.

\vspace*{0.7cm}
\noindent
{\bf Table 4:} Cross sections and event fractions for missing
energy plus jets events for various SUSY cases
at the LHC.

\begin{center}
\begin{tabular}{|c|cccccccc|}
\hline
case & $\sigma (pb)$ & $n_j=4-5$ & $6-7$ & $\ge 8$ &
$n_b\ge 1$ & $n_b\ge 2$ & $\langle \Sigma E_T\rangle $ & $\langle \eslt
\rangle $ \\
\hline
\hline
A1 & 24.6 & 0.54 & 0.35 & 0.10 & 0.65 & 0.28 & 1160 & 212 \\
A2 & 48.0 & 0.55 & 0.35 & 0.10 & 0.47 & 0.09 & 1112 & 221 \\
B1 & 79.1 & 0.73 & 0.25 & 0.02 & 0.21 & 0.04 & 964 & 195 \\
B2 & 67.7 & 0.77 & 0.20 & 0.03 & 0.23 & 0.05 & 999 & 211 \\
B3 & 51.8 & 0.57 & 0.35 & 0.08 & 0.36 & 0.12 & 1118 & 215 \\
B4 & 20.1 & 0.54 & 0.36 & 0.10 & 0.44 & 0.15 & 1204 & 217 \\
BTW1 & 105 & 0.69 & 0.26 & 0.04 & 0.17 & 0.04 & 1006 & 214 \\
BTW2 & 57.3 & 0.61 & 0.33 & 0.05 & 0.18 & 0.04 & 1091 & 211 \\
BTW3 & 96 & 0.69 & 0.27 & 0.04 & 0.15 & 0.03 & 1011 & 217 \\
BTW4 & 52.3 & 0.60 & 0.33 & 0.06 & 0.18 & 0.04 & 1109 & 208 \\
$t\bar t(160)$ & 2.9 & 0.81 & 0.17 & 0.01 & 0.56 & 0.12 & 895 & 201 \\
$Z+jets$ & 0.63 & 0.89 & 0.11 & 0.0 & 0.11 & 0.02 & 905 & 281 \\
\hline
\end{tabular}
\end{center}
\vspace*{0.7cm}

The most useful discriminators appear to be the fractions of events with at
least one or two tagged $b$ quarks. Here we have assumed a $b$ tagging
efficiency of 40\% if the $b-$flavoured hadron has $p_T > 20$ GeV and
pseudorapidity $|\eta| < 2$, and zero efficiency otherwise; we have ignored
the possibility of false tags. The spectra where $\tilde{g} \rightarrow \st +
t$ is allowed clearly lead to the largest $b$ content; the difference between
$\st \rightarrow \tw_1 + b$ (A1) and $\st \rightarrow \lsp + c$ (A2) decays is
also evident, especially in the fraction of events with at least two tagged
$b$'s. Cases B3 and B4 still have fairly large $b$ content, partly due to
$\tilde{g} \rightarrow \tw_1 t b$ decays which can produce up to four $b$
quarks in a gluino pair event; enhanced $\tilde{g} \rightarrow \tilde{Z}_2 +
\bbbar$ and $\tilde{Z}_2 \rightarrow \lsp + \bbbar$ branching ratios also play
a role, see table 2. Finally, in the light gluino scenarios B1, B2 the
$b-$fraction is considerably smaller than in the cases where gluinos can decay
into top quarks; the enhanced branching ratios into final states containing
\bbbar\ still lead to $b-$fractions that exceed those of the BTW cases,
however.

The presence of $t$ quarks in gluino decays also leads to large average jet
multiplicities: 10\% of all events in cases A1, A2 and B4 have at least 8
reconstructed jets in them (recall that we require each jet of have $p_T > 50$
GeV). Moreover, the comparison of BTW2,4 with BTW1,3 shows that the presence
of first generation squarks significantly, but not infinitely, heavier than
the gluino increases the average number of jets and the average scalar $E_T$
per event; this is due to $\tilde{g} \tilde{q}$ production, of
course. This effect is less pronounced when comparing B4 to B3, or B2 to B1,
since in cases B4 and B2 first generation squarks are so heavy that they
contribute little to the total SUSY signal; see table 2. The average missing
$E_T$ seems to be a less useful discriminator, clustering around 210 to 220
GeV in almost all cases. The only exception is case B1, which has a very light
gluino; in case B2, where the gluino is also light, the missing $E_T$ is
enhanced by the large branching ratios for loop--induced $\tilde{g}
\rightarrow \tilde{Z}_{1,2} + g$ decays discussed earlier.

Similar tables can be shown for the other samples of signal events \cite{16};
for reasons of space I here merely summarize the most important points. The
results for jet multiplicities are very similar for all samples, except that a
larger number of leptons implies an overall reduction of the number of
jets.\footnote{Recall that our definition of leptonic events does not require
a minimal number of jets; this also reduces the average jet multiplicity
compared to the missing $E_T$ sample, of course.} The $b-$content of events in
the $1l$ sample are similar to those in the missing $E_T$ sample, with a
slight enhancement in case B1 and slight reductions in the BTW cases which
increases the differences between these scenarios.

In the mSUGRA cases B1--B4 the $b$ fraction is considerably larger in the OS
sample than in the missing $E_T$ or $1l$ samples. This effect is especially
pronounced in cases B3 and B4, where $\tilde{g} \rightarrow \tw_1 t b$ decays
are good sources of both hard leptons and $b$ quarks; indeed, in these two
cases the $b$ fraction exceeds that in case A2 in the OS dilepton sample. The
correlation between the number of $b$'s and the number of leptons in cases B1
and B2 is due to the fact that here most leptons come from $\tilde{Z}_2$
decays, and $\tilde{g} \rightarrow \tilde{Z}_2$ decays frequently lead to a
\bbbar\ final state; the enhancement of \bbbar\ final states is much weaker in
$\tilde{g} \rightarrow \lsp$ decays, as shown in table 1, while $\tilde{g}
\rightarrow \tw_1$ decays cannot contain $b$ quarks in these cases.

The contribution from $\tilde{Z}_2 \rightarrow \lsp l^+ l^-$ decays to the OS
signal can also be tested by looking at the flavour of the produced leptons:
$\tilde{Z}_2$ decays always produce \eplem\ or $\mu^+ \mu^-$ pairs, while OS
events that originate from (semi--)leptonic decays of charginos or $t$ quarks
are equally likely to contain an $e \mu$ pair. Indeed, one finds a strong
preponderance of like--flavour pairs in cases B1 and B2 as well as in all BTW
cases, while all combinations of lepton flavours occur with equal frequency in
cases A1 and A2 where $\tilde{Z}_2$ is hardly ever produced; in cases B3 and
B4 there is a smaller but still significant preference for like--flavour
lepton pairs. The invariant mass distribution of the two charged leptons in
like--flavour opposite--charge dilepton events should allow to determine the
$\tilde{Z}_2 - \lsp$ mass difference, which in our mSUGRA scenario is
approximately given by $0.4 \mhalf$. Unfortunately, other (combinations of)
sparticle masses are much more difficult to determine at hadron colliders.

\section*{5) Summary and Conclusions}
In this contribution I have described the phenomenology of the minimal SUGRA
$SU(5)$ model. The underlying theme of this model is unification. First of
all, the very existence of a GUT sector implies that the Higgs sector of the
SM is ill--behaved at the quantum level unless SUSY exists at an energy scale
not much above the weak scale. Moreover, we now know that the particle content
of the SM by itself does not allow for unification of all gauge interactions;
new degrees of freedom are needed, and minimal SUSY just fits the bill.

Secondly, in this model one can also successfully unify the $b$ and $\tau$
Yukawa couplings, provided that the top Yukawa coupling is large, i.e. close
to its IR (quasi) fixed point. The idea of unification can even be extended
into the SUSY breaking sector by assuming a single gaugino mass as well as a
single scalar mass. This leads to the very elegant mechanism of radiative
breaking of the electroweak gauge symmetry, which hints towards explanations
for the large mass of the top quark and the large hierarchy between \mx\ and
\mz.

In this scenario constraints from proton decay, and from the relic density of
LSPs produced in the Big Bang, imply that gauginos must be quite light, while
most sfermions and Higgs bosons are heavy. Light charginos, and the light
scalar Higgs boson predicted by this model, are quite easily detectable at
\eplem\ colliders like the second stage of LEP if kinematically accessible. If
no chargino with mass below $\sim 105$ GeV or no gluino with mass below $\sim
400$ GeV is found mSUGRA $SU(5)$ has to be discarded, but their discovery can
hardly be considered proof of this model. However, it also predicts that
left--handed $\tilde{b}$ squarks and, much more dramatically, the lighter
$\tilde{t}$ eigenstate lie well below the other squarks. Indeed, \st\ might
even be produced in $\tilde{g}$ decays, giving events with several $b-$quarks.
Even if this decay is not possible the reduced masses of \st\ and
$\tilde{b}_L$ give enhanced branching ratios for $\tilde{g} \rightarrow \tw_1
t b$ and $\tilde{g} \rightarrow \tilde{Z}_{1,2} \bbbar$ three--body decays and
hence again an enhaced $b-$quark content of SUSY events at the LHC; in this
case the enhancement is usually less, and increases with the number of hard
leptons in the event. Notice that the event rates are so large that a ``low"
luminosity of a few times $10^{32}$ cm$^{-2}$sec$^{-1}$ is quite sufficient,
which should allow for $b-$tagging at least in principle. Further clues to the
nature of the spectrum can come, e.g., from the average jet mulitplicity per
event and from the flavour composition of opposite--sign dilepton events.

It should be emphasized that in this model a SUSY signal should be seen at the
LHC in at least three, and possibly as many as seven independent channels,
where only the number and charge of hard leptons has been used to classify
events. This clearly offers great opportunities for the LHC. Nevertheless the
LHC by itself will not suffice to find all the new particles predicted by
this model: The heavy Higgs bosons, higgsinos and sleptons are in my opinion
impossible to detect at the LHC. Finding first generation squarks on top of
the ``background" of light gluinos will also be difficult, especially if they
are close to the TeV scale in mass. More surprisingly, even the direct pair
production of light \st\ squarks seems difficult to detect. If we are lucky
light stops might be detected at the tevatron \cite{38}. However, if this
model is correct the completion of the sparticle and Higgs spectrum will most
likely have to await the construction of a TeV scale \eplem\ collider.
Finally, one would eventually want to see direct evidence for the existence of
GUT particles, the best hope probably being proton decay experiments. The
model presented here would therefore keep particle physicists busy for some
decades to come.

\subsection*{Acknowledgements}
I thank my collaborators Howie Baer, Chung Kao, Mihoko Nojiri and Xerxes Tata,
who did all the work while I got to swim in the Indian ocean. I also thank the
workshop organizers for their kind invitation, which helped me to complete
some biochemical research as well. This work was supported in part by the U.S.
Department of Energy under contract No. DE-AC02-76ER00881, by the Wisconsin
Research Committee with funds granted by the Wisconsin Alumni Research
Foundation, as well as by a grant from the Deutsche Forschungsgemeinschaft
under the Heisenberg program.


\newpage
\section*{Figure Captions}
\renewcommand{\labelenumi} {Fig. \arabic{enumi}}
\begin{enumerate}

\item   
The ratio $m_{\tilde{b}_1}/m_{\tilde{u}_L}$ as a function of the GUT scale
parameter $A_0 \equiv A/m_0$.

\vspace{6mm}
\item   
Regions in the $(A_0,\mhalf)$ plane leading to different gluino and \st\
decays. Note that $m_{\tilde{t}_1}$ generally increases with increasing
\mhalf\ and decreasing $|A_0|$, although for given \mhalf\ the maximum of
$m_{\tilde{t}_1}$ is not exactly at $A_0=0$.

\end{enumerate}
\end{document}